\documentclass{moriond}

\usepackage{amsthm}
\usepackage{amsmath}
\usepackage{amssymb}

\def\be{\begin{equation}}
\def\ee{\end{equation}}
\def\bea{\begin{eqnarray}}
\def\eea{\end{eqnarray}}

\newcommand{\Photo}{}

\begin{document}
\vspace*{4cm}
\title{Pheno \& Cosmo Implications of Scotogenic 3-loop Neutrino Mass Models}

\author{Asmaa~Abada $^a$, Nicol\'{a}s Bernal $^b$, Antonio E. C\'{a}rcamo Hern\'{a}ndez $^{c,d,e}$, Sergey Kovalenko $^{d,e,f}$, $^*$T\'{e}ssio B. de Melo $^{e,f}$, Takashi Toma $^{g,h}$}

\address{$^a$ P\^ole Th\'eorie, Laboratoire de Physique des 2 Infinis Ir\`ene Joliot Curie (UMR 9012)\\
CNRS/IN2P3, 15 Rue Georges Clemenceau, 91400 Orsay, France \\
$^b$New York University Abu Dhabi, PO Box 129188, Saadiyat Island, Abu Dhabi, United Arab Emirates \\
$^c$Universidad T\'{e}cnica Federico Santa Mar\'{\i}a, Casilla 110-V, Valpara\'{\i}so, Chile \\
$^d$Centro Cient\'{\i}fico-Tecnol\'{o}gico de Valpara\'{\i}so, Casilla 110-V, Valpara\'{\i}so, Chile \\
$^e$Millennium Institute for Subatomic Physics at the High-Energy Frontier, SAPHIR, Chile \\
$^f$Universidad Andr\'es Bello, Facultad de Ciencias Exactas, \\ Departamento de Ciencias Físicas-Center for Theoretical and Experimental Particle Physics, \\
Fernández Concha 700, Santiago, Chile\\
$^g$Institute of Liberal Arts and Science, Kanazawa University, Kanazawa 920-1192, Japan \\
$^h$Institute for Theoretical Physics, Kanazawa University, Kanazawa 920-1192, Japan}

\maketitle\abstracts{
Radiative seesaw models are examples of interesting and testable extensions of the Standard Model to explain the light neutrino masses. In radiative models at 1-loop level, such as the popular scotogenic model, in order to successfully reproduce neutrino masses and mixing, one has to rely either on unnaturally small Yukawa couplings or on a very small mass splitting between the CP-even and CP-odd components of the neutral scalar mediators. We discuss here scotogenic-like models where light-active neutrino masses arise at the three-loop level, providing a more natural explanation for their smallness. The proposed models are consistent with the neutrino oscillation data and allow to successfully accommodate the measured dark matter relic abundance. Depending on the specific realization, it is also possible to explain the W-mass anomaly and to generate the baryon asymmetry of the Universe via leptogenesis. The models lead to rich phenomenology, predicting sizable charged-lepton flavor violation rates, potentially observable in near future experiments, while satisfying all current constraints imposed by neutrinoless double-beta decay, charged-lepton flavor violation and electroweak precision observables.}

\section{Introduction}

The simplest and most elegant extensions of the Standard Model (SM) that provide masses for the light active neutrinos employ the seesaw mechanism, which relies on adding new particles to the SM content while keeping its gauge symmetry, restulting in neutrino masses generated at tree level. To comply with neutrino oscillation data, the additional states are either too heavy to be detected or interact with the SM via tiny Yukawa couplings. In both cases, the possibilities of testing the mass generation mechanism are very limited. Alternatively, radiative seesaw models are viable and testable extensions of the SM explaining the tiny neutrino masses and their mixings, while the seesaw mediators play an important role in successfully accommodating the observed amount of Dark Matter (DM) relic density~\cite{Cai:2017jrq,Jana:2019mgj,Arbelaez:2022ejo,Bonilla:2016diq,Baek:2017qos}. In most radiative seesaw models, light neutrino masses are generated at the 1-loop level. However, this framework still requires very small Yukawa couplings, of the order of ${\mathcal{O}}(10^{-6})$, or unnaturally small mass splitting between the CP-even and CP-odd components of the neutral scalar mediators.

In this work, we explore models where light-active neutrino masses arise at the 3-loop level, providing a more natural explanation for the smallness of the neutrino masses, and also accomodating viable DM candidates. The first model, discussed in Section~\ref{Sec:model_1}, is an extended scotogenic model~\cite{Tao:1996vb,Ma:2006km}, while the second, discussed in Section~\ref{Sec:model_2}, has an inverse seesaw (ISS) structure~\cite{Mohapatra:1986bd}. After presenting a description of the field content and symmetries, we examine their key phenomenological aspects, including charged-lepton flavor violation (cLFV) and electroweak precision observables. Additionally, we show that besides accomodating neutrino masses and DM, the first model can account for the W-mass anomaly, while the second model can explain the Baryon Asymmetry of the Universe (BAU) via leptogenesis. We summarize our findings in Section~\ref{Sec:conclusions}.

\section{Scotogenic 3-loop Model}
\label{Sec:model_1}

\begin{figure}[!t]
    \begin{center}
        \includegraphics[width=0.45\textwidth]{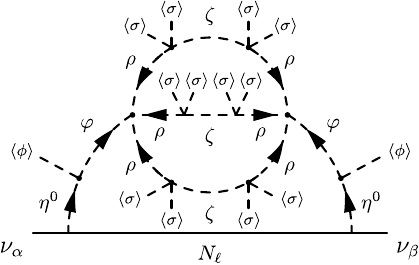} 
    \end{center}
    \caption{Scotogenic 3-loop diagram responsible for neutrino masses, with $\ell=1$, 2 and $\protect\alpha,\, \protect\beta = e,\, \protect\mu,\, \protect\tau$.}
    \label{Neutrinodiagram}
\end{figure}

We first discuss a 3-loop neutrino mass model which resembles more closely the original scotogenic model~\cite{Tao:1996vb,Ma:2006km}. It has an augmented symmetry group, encompassing a spontaneously broken global symmetry $U(1)'$ and a preserved discrete symmetry $\mathbb{Z}_2$.
The new particles consist of an inert scalar doublet $\eta$ and two RH neutrinos $N_{R_k}$, which have non-trivial $\mathbb{Z}_2$ parities. 
In addition, four electrically neutral scalar singlets $\sigma$, $\rho$, $\varphi$, $\zeta$ are also included, all of them odd under $\mathbb{Z}_2$, except for $\sigma$, which is responsible for breaking the $U(1)'$ symmetry at the TeV scale. 
The Yukawa interactions, consistent with these symmetries and particle content, are given by
\begin{align}
    -\mathcal{L}_Y \supset y_{u \phi}^{ij}\, \bar{q}_{iL} \widetilde{\phi} u_{jR} + y_{d \phi}^{ij}\, \bar{q}_{iL} \phi d_{jR} + y_{l \phi}^{ij}\, \bar{\ell}_{iL} \phi \ell_{R_j} + y_\eta^{ik}\, \bar{\ell}_{iL} \widetilde{\eta} N_{R_{k}} + M_{N_{R}}^{kr}\, \bar{N}_{R_{k}} N_{R_{r}}^C + \mathrm{H.c.}
\label{eq:lagrangian2}\ .\end{align}
It is worth noting that neutrino masses vanish at the tree level, due to the absence of a Dirac mass term connecting the left and right-handed neutrinos after spontaneous symmetry breaking. Additionally, the symmetries imposed in the model preclude both 1-loop and 2-loop-level contributions. However, through the Yukawa interactions and the scalar potential terms,
\begin{align}
    V& \supset \lambda _{15}\left( \rho \zeta \sigma ^{2}+\mathrm{H.c.}\right) + \lambda_{14}\left( \varphi \rho ^{3}+\mathrm{H.c.}\right) + A\left[ (\eta ^{\dagger }\phi )\varphi +\mathrm{H.c.}\right] ,
    \label{eq:sacalar-potential}  
\end{align}
a 3-loop contribution is allowed, as depicted in Fig. \ref{Neutrinodiagram}. This 3-loop diagram provides the leading contribution to neutrino masses in this setup.

The $\mathbb{Z}_2$ symmetry protects the lightest odd state, making it a viable candidate for DM. In the present scenario, one could have a scalar ($\eta^0$, $\varphi$, $\rho$ or $\zeta$) or fermionic ($N_{R_1}$) DM, depending on the mass hierarchy. 
The case of scalar DM corresponds to the scenario in which the candidate for DM is the lightest of $\eta^0$, $\varphi$, $\rho$, and $\zeta$. 
If the dominant component of DM is $\eta^0$, the properties of DM are similar to those of the inert scalar DM~\cite{LopezHonorez:2006gr}. 
The main annihilation channels are into the gauge bosons and Higgs bosons via the gauge interactions. 
On the other hand, if DM is dominated by the other components, the main annihilation channels are into the Higgs bosons via the scalar couplings. This scenario corresponds to the Higgs portal DM~\cite{McDonald:1993ex,Burgess:2000yq}. 
For fermionic DM, the candidate is the lightest state $N_{R_1}$. In the early Universe, it can be annihilated into a couple of charged leptons or active neutrinos via the $\eta$ exchange ($t$- and $u$-channels). The DM properties for this scenario are similar to those of the original scotogenic model~\cite{Ma:2006km,Kubo:2006yx,Cacciapaglia:2020psm,Rosenlyst:2021tdr}.

\paragraph{Oblique parameters and $W$ boson mass}

\begin{figure}[t]
    \centering
    \includegraphics[width=0.445\linewidth]{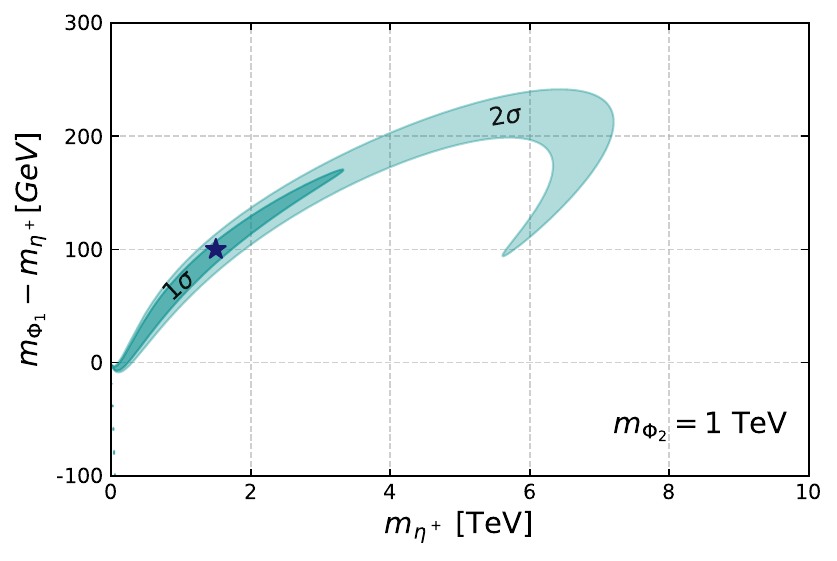}
    \includegraphics[width=0.525\linewidth]{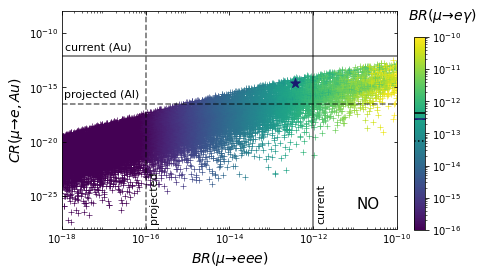}
    \caption{\textit{Left panel}: the $1$-$\protect\sigma$ and $2$-$\protect\sigma$ regions in the $m_{\Phi_1} - m_{\protect\eta^+}$ versus $m_{\protect\eta^+}$ plane to account for the CDF measurement of the $W$ mass. \textit{Right panel}: correlation among the cLFV processes $\mu \to eee$ and $\mu \to e$ conversion in gold nuclei for normal neutrino mass ordering. The points are colored according to the size of the branching ratio of the $\mu \to e \gamma$ process, as shown in the side bar. The current upper bounds are indicated by the black full lines, while the future sensitivities, by the black dashed lines.}
    \label{stuCDFplot}
\end{figure}

The measurement of the $W$ gauge boson mass by the CDF collaboration~\cite{CDF:2022hxs} can be interpreted as new physics contributions to the oblique corrections, usually parameterized in terms of the well-known quantities $S$, $T$ and $U$: 
\begin{equation}
M _W ^2 = \left( M _W ^2 \right) _\text{SM} + \frac{\alpha _\text{EM} \left( M _Z \right) \cos ^2 \theta _W \, M _Z ^2}{\cos ^2 \theta _W - \sin ^2 \theta _W} \left[ - \frac{S}{2} + \cos ^2 \theta _W \, T + \frac{\cos ^2 \theta _W - \sin ^2 \theta _W}{4\, \sin ^2 \theta _W}\, U \right] .
\end{equation}
The presence of extra scalars in our model, in particular those belonging to the inert doublet, affect the oblique corrections of the SM. In other words, the CDF anomaly can be explained by non-trivial $S$, $T$, and $U$ values generated by the extra scalars. The CDF measurement can be explained in this model with scalar masses in the TeV scale, as long as that the mass splitting among the charged and neutral scalars is within a few hundred GeV. In the left panel of Fig.~\ref{stuCDFplot}, we show the $1$-$\protect\sigma$ and $2$-$\protect\sigma$ parameter space regions in which the CDF anomaly can be accomodated.

\paragraph{Charged lepton flavor violating observables}

The cLFV radiative decays $\mu \rightarrow e\gamma $ and $\mu \to e e e $, and $\mu-e$ conversion in atomic nuclei, are excellent probes to test this model. They arise at 1-loop level from the exchange of charged scalars $\eta ^\pm$ and RH neutrinos $N_{R_k}$.
We perform a random scan over the parameters of the model and compute the corresponding cLFV rates. 
The result is shown in the right panel of Fig.~\ref{stuCDFplot}.
We can see that a large portion of the parameter space that pass the current constraints will be probed in future $\mu-e$ conversion and $\mu \to e e e$ experiments. In Fig.~\ref{stuCDFplot} we also show a benchmark point, marked as a blue star, which is an instance of a point that comply with all the current constraints, accomodates neutrino masses and DM, account for the $W$ mass anomaly and is within the reach of future cLFV experiments. 

\section{3-loop Inverse Seesaw Model}
\label{Sec:model_2}

\begin{figure}[!t]
    \begin{center}
        \includegraphics[width=0.45\textwidth]{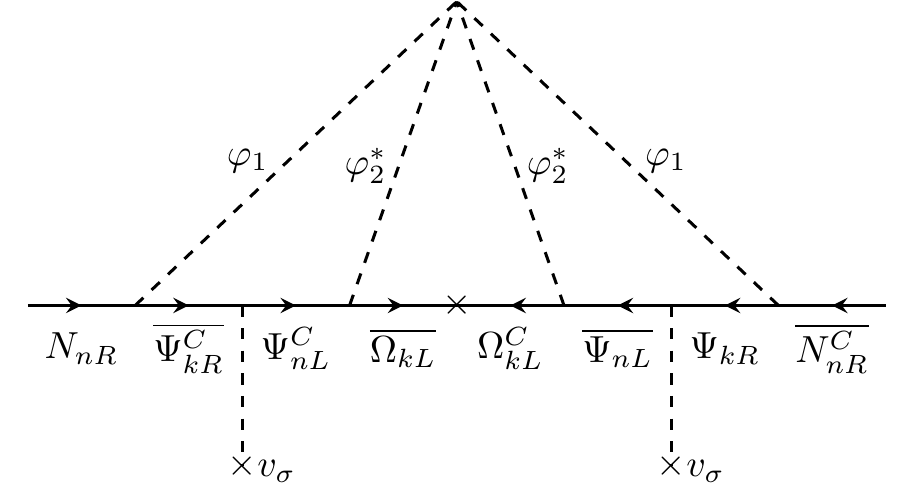} 
    \end{center}
    \caption{Three-loop diagram for the lepton number violating Majorana mass in the ISS, with $n, k = 1, 2$.}
    \label{diagram_inv}
\end{figure}

In this section, we present a second instance of a 3-loop neutrino mass model, this time featuring an inverse seesaw structure. This is an extension of the SM with gauge singlet fields: three scalars $\varphi_{1,2}, \sigma$, two left-handed Majorana neutrinos $\Omega_{1,2}$ and two vector-like neutral leptons $\Psi_{1,2}$. The SM gauge symmetry is extended with the global symmetry $U(1)' \otimes \mathbb{Z}_2$. The global $U(1)'$ symmetry is spontaneously broken at the TeV scale by the VEV of $\langle\sigma\rangle$ down to a residual preserved $\mathbb{Z}_4$ symmetry. This results in the new particles acquiring masses on the TeV scale. The Yukawa interactions relevant to the neutrino mass generation in this setup are given by 
\begin{align} \label{Eq:Lag3}
- & \mathcal{L} _Y^{\left(\nu\right)} = \sum_{i=1}^3 \sum_{k=1}^2 \left(y_{\nu}\right)_{ik}\, \overline{l}_{iL}\, \widetilde{\phi}\, \nu_{kR} + \sum_{n, k=1}^2\, M_{nk}\, \overline{\nu}_{nR}\, N_{kR}^C + \sum_{n, k=1}^2 \left(y_{N}\right)_{nk}N_{nR}\, \varphi_1^*\, \overline{\Psi_{kR}^C} \notag \\
&+ \sum_{n, k=1}^2 \left(y_{\Omega}\right)_{nk}\, \Psi_{nL}^C\, \varphi_2\overline{\Omega}_{kL} + \sum_{n, k=1}^2 \left(y_{\Psi}\right)_{nk}\, \overline{\Psi}_{nL}\, \sigma\, \Psi_{kR} + \sum_{n=1, k}^2 \left(m_{\Omega}\right)_{nk}\, \overline{\Omega}_{kL}\, \Omega_{nL}^C + \text{H.c.}
\end{align}
The scalar fields $\varphi_1$ and $\varphi_2$ do not acquire vacuum expectation values, but together with the heavy neutral leptons $\Psi_{kR}$, $\Psi_{kL}$ and $\Omega_{kL}$ (with $k=1$, 2) induce the LNV Majorana mass term $\mu_{nk}\, \overline{N_{nR}}\, N_{kR}^C$ at the three-loop level according to the diagram shown in Fig.~\ref{diagram_inv}. The obtained physical neutrino spectrum exhibits the typical spectrum pattern of the inverse seesaw mechanism. It is composed of three light active neutrinos and four exotic neutral states, forming two pairs of pseudo-Dirac neutrinos, denoted as $N ^\pm _k$. The small mass splitting of the quasi-degenerate pseudo-Dirac pairs is proportional to the small Majorana mass scale $\mu $.

\paragraph{Dark Matter, Leptogenesis and Charged lepton flavor violation}

Depending on the mass hierarchy, we could have scalar or fermionic DM candidates. We focus on the scenario in which the lightest odd state is $\Psi _{1R}$, so we have a fermionic DM candidate. This case is particularly interesting because the Yukawa coupling that mediates DM annihilation also participates in the calculation of the $\mu$ parameter, allowing us to correlate the cFLV, DM, and neutrino masses. For DM masses in the GeV-to-TeV ballpark and Yukawa couplings at the electroweak scale, DM could have been generated in the early Universe via the WIMP mechanism. The baryon asymmetry of the Universe can also be generated in this model via leptogenesis, in the presence of out-of-equilibrium CP and lepton number violating decays of $N_k^{\pm}$. We assume that $| M_{N _1 ^\pm} | \ll | M_{N _2 ^\pm} | $ so that only the first generation of $N_k^{\pm}$ contribute to the BAU. The correct lepton asymmetry parameter can be obtained both in the strong and weak washout regimes thanks to the small sppliting between $N _1 ^+$ and $N _1 ^-$, proportional to the 3-loop generated $\mu$.

\begin{figure}[ht]
    \begin{center}
    \includegraphics[width=0.475\linewidth]{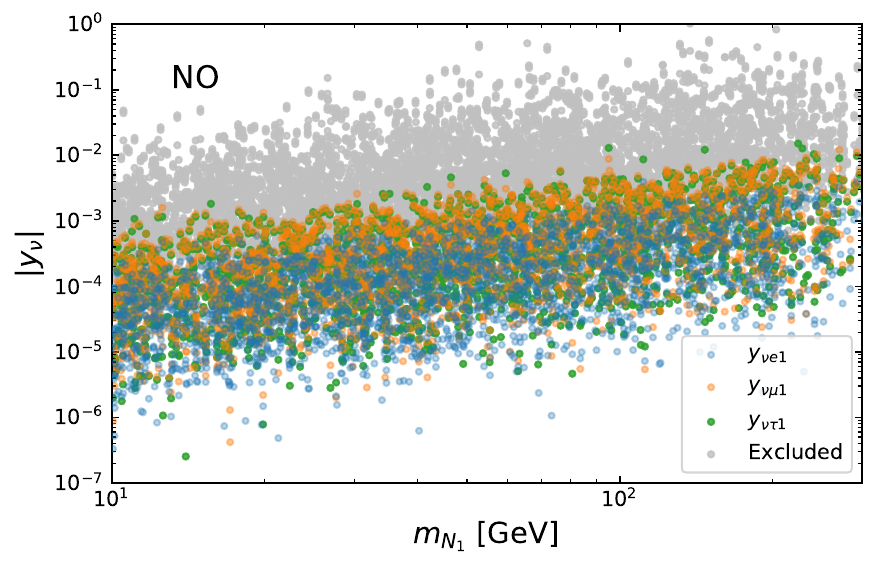}
    \includegraphics[width=0.47\linewidth]{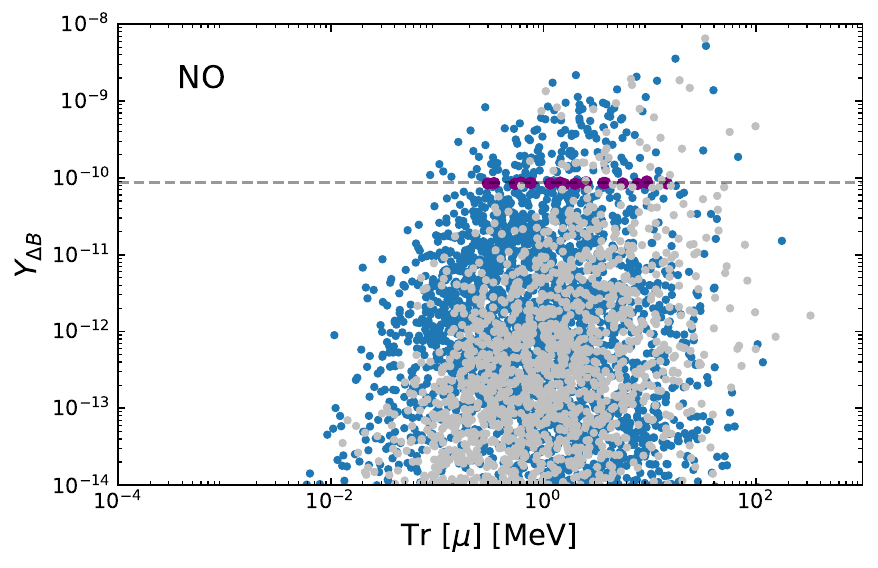}
    \end{center}
    \caption{\textit{Left panel}: allowed values in the $|y_{\nu}|$-$m_{N_1}$ plane consistent with the DM relic abundance and neutrino oscillation data assuming normal ordering. The gray points are excluded by cLFV constraints. \textit{Right panel}: baryon asymmetry parameter $Y_{\Delta B}$ as a function of the trace of the Majorana matrix $\mu$. All points comply with the DM relic abundance and neutrino oscillation data assuming normal ordering. The purple points have the correct $Y_{\Delta B}$ within $3 \sigma$, while the gray points are excluded by cLFV. The values of $\mu$ in the keV-MeV range are favored by leptogenesis.}
    \label{yvsmN}
\end{figure}

\begin{figure}[b]
    \centering
    \includegraphics[width=0.505\linewidth]{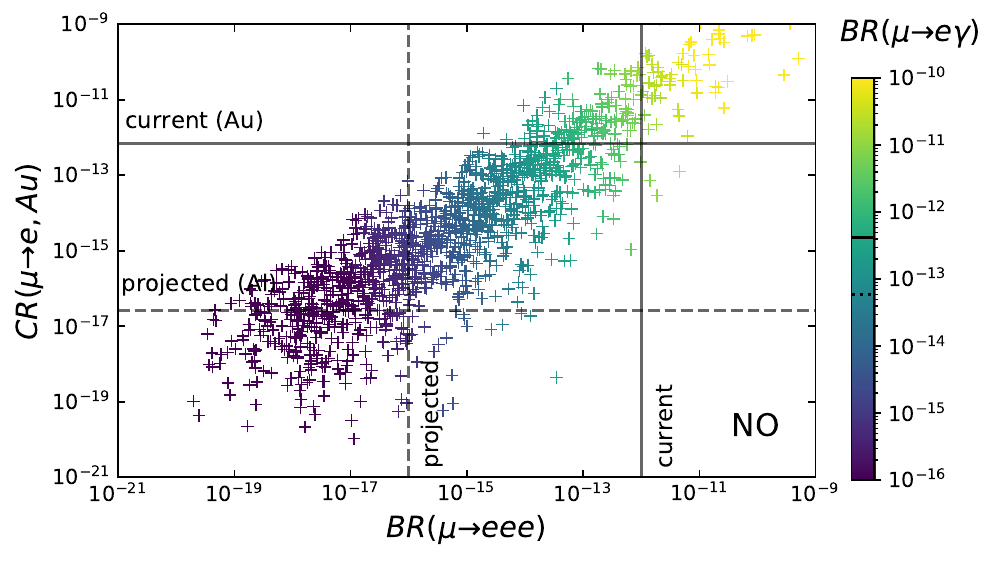}
    \caption{Correlation among cLFV observables. Points comply with the BAU, DM relic abundance and neutrino oscillation data assuming normal ordering. Current bounds are shown as full lines, while projections are shown as dashed lines.}
    \label{cLFV_2}
\end{figure}

We show in the left panel of Fig.~\ref{yvsmN} the allowed parameter space in the $|y_{\nu}|$-$m_{N_1}$ plane consistent with the constraints imposed by the measured value of the DM relic abundance and by the experimental neutrino oscillation data. The consistency with the constraints arising from the cLFV requires Yukawa couplings $y _{\nu _{ik}}$ smaller than $10^{-2}$. The points excluded by cLFV constraints are the gray points in Fig.~\ref{yvsmN}. 
In the right panel of Fig.~\ref{yvsmN} we show the baryon asymmetry parameter $Y_{\Delta B}$ as a function of the trace of the Majorana matrix $\mu$. The gray points in this plot again represent excluded points by cLFV constraints. The purple points correspond to the values of $Y_{\Delta B}$ within the experimentally allowed range at $3 \sigma$~CL, which is possible provided that the entries of the Majorana submatrix $\mu$ are in the keV to MeV range. Finally in Fig.~\ref{cLFV_2} we present the correlation among the cLFV rates with the corrent and future constraints, showing that this model also has a significant potential of producing signals in future cLFV experiments.

\section{Conclusion} \label{Sec:conclusions}

We have examined two models where the tiny masses of the active neutrinos are radiatively generated at three-loop level. In these setups, the 3-loop suppression allows the new particles to have masses in the TeV scale without fine-tuning the Yukawa couplings. 
We found that these models successfully complies with the constraints imposed by the neutrino oscillation experimental data, neutrinoless double beta decay, dark matter relic density, charged lepton flavor violation, electron-muon conversion. In one of the models, reminiscent of the original scotogenic model, the W-mass anomaly finds a simple explanation, while the other model, which has an inverse seesaw structure, provides essential means for efficient low-scale resonant leptogenesis. We have also shown that in the models considered, the charged lepton flavor violating decays $\mu\to e\gamma$, $\mu\to eee$ as well as the electron-muon conversion processes get sizable rates, which are within the reach of sensitivity of the forthcoming experiments.

\section*{Acknowledgments}

\noindent
The results here presented are based on Refs.~\cite{Abada:2022dvm,Abada:2023zbb}. 
TBM thank the organizers for the invitation to speak at the Rencontres de Moriond EW 2024.
NB received funding from the Spanish FEDER/MCIU-AEI under grant FPA2017-84543-P. 
This project has received funding and support from the European Union's Horizon 2020 research and innovation programme under the Marie Sk{\l}odowska-Curie grant agreement
No.~860881 (H2020-MSCA-ITN-2019 HIDDeN) and from the Marie Sk{\l}odowska-Curie Staff Exchange grant agreement No 101086085 ``ASYMMETRY''. A.E.C.H.. and S.K. are supported
by ANID-Chile FONDECYT 1170171, 1210378, 1230160, ANID PIA/APOYO AFB230003, and Proyecto Milenio- ANID: ICN2019\_044. TBM acknowledges ANID-Chile grant FONDECYT No. 3220454 for fnancial support.
This work was supported by the JSPS Grant-in-Aid for Scientific Research KAKENHI Grant No. JP20K22349 (TT).

\section*{References}

\bibliographystyle{unsrt}

\end{document}